\documentclass[3p,times]{elsarticle}
\usepackage{graphicx} % Include figure files
\usepackage{dcolumn}  % Align table columns on decimal point
\usepackage{color}
\usepackage{wasysym}

\newcommand{\kl}   {\ensuremath{K_L^{\,0}}}
\newcommand{\gevc} {\ensuremath{\,{\mathrm{GeV}/c}}}

\newcommand{\mev}  {\ensuremath{\,{\mathrm{MeV}}}}
\newcommand{\hvb}  {\ensuremath{{V_{\mathrm{break~down}}}}}
\journal{Nuclear Instruments and Methods in Physics Research Section A}

\begin{document}

\begin{frontmatter}

\title{A scintillator based endcap $K_L$ and muon detector for the
  Belle II experiment}

\author[mephi,itep,mipt]{T.~Aushev}
\author[mephi,kansas]{D.\,Z.~Besson}
\author[mephi,itep]{K.~Chilikin}
\author[mephi,itep]{R.~Chistov}
\author[mephi,itep]{M.~Danilov}
\author[mephi,itep]{P.~Katrenko}
\author[mephi,itep]{R.~Mizuk}
\author[mephi,itep]{G.~Pakhlova}
\author[mephi,itep]{P.~Pakhlov}
\author[mephi,itep]{V.~Rusinov}
\author[mephi,itep]{E.~Solovieva}
\author[mephi,itep]{E.~Tarkovsky}
\author[mephi,itep]{I.~Tikhomirov}
\author[mephi,itep,mipt]{T.~Uglov}

\address[mephi]{National Research Nuclear University MEPhI (Moscow Engineering Physics
Institute), \\
Kashirskoye shosse 31, Moscow, 115409, Russia \\}

\address[itep]{Institute for Theoretical and Experimental Physics, \\
B. Cheremushkinskaya 25, Moscow, 117218, Russia \\}

\address[mipt]{Moscow Institute of Physics and Technology, \\
Institutskiy per. 9, Dolgoprudny, Moscow Region, 141700, Russia \\}

\address[kansas]{University of Kansas, Department of Physics and Astronomy, \\
1082 Malott Hall, Lawrence, KS 66045, USA \\}

\begin{abstract}
  A new \kl\ and muon detector based on scintillators will be used for
  the endcap regions in the Belle II experiment, currently under
  construction. The increased luminosity of the $e^+ e^-$ SuperKEKB
  collider entails challenging detector requirements. We demonstrate
  that relatively inexpensive polystyrene scintillator strips with
  wavelength shifting fibers ensure a sufficient light yield at the
  Silicon PhotoMultiplier (SiPM) photodetector, are robust and provide
  improved physics performance for the Belle II experiment compared to
  its predecessor, Belle.
\end{abstract}

\begin{keyword}
Plastic scintillator detectors \sep
Particle tracking detectors\sep  
SiPM\sep 
Muon detector
\end{keyword}
\end{frontmatter}

\section{Overview}

Over the last decade, the B-factory experiments, Belle and BaBar, have
shown that flavor physics has the powerful potential to search for
various manifestations of New Physics. If the statistical errors of
measurements in the flavor sector can be substantially improved, the
energy scale of New Physics studies can be pushed beyond 1~TeV,
providing strong impetus for construction of the next generation
B-factory. The idea of an upgraded Belle experiment was first
presented in a Letter of Intent in 2004~\cite{LoI}, followed by a
Technical Design Report in 2010~\cite{belle2tdr}. In parallel, the
KEKB accelerator group has defined the parameters of the SuperKEKB
accelerator, an upgraded version of KEKB, with luminosity increased by
almost two orders of magnitude, and an ultimate instantaneous
luminosity goal of $8 \times 10^{35}~\mathrm{cm}^{-2}
\mathrm{s}^{-1}$. Currently the SuperKEKB Factory, and the associated
Belle II sub-detectors are under construction at the High Energy
Accelerator Research Organization, KEK, in Tsukuba, Japan. This new
installation is expected to begin operation at the end of 2015 with
the goal of collecting an integrated luminosity of 50~ab$^{-1}$ by
2020.

The \kl\ and muon subsystems (KLM) in the Belle experiment were
designed to detect \kl\ mesons and muons as they traversed the
segmented flux return of the Belle solenoid, using resistive plate
chambers (RPC)~\cite{belle,RPC}. The KLM system has a cylindrical
(barrel) part and two planar endcap sections. The RPC KLM detector
operated successfully over the entire lifetime of Belle (1999-2010).
However, the endcap KLM gas detectors of the Belle II experiment would
suffer considerably compromised performance in the higher luminosity
SuperKEKB environment, given the higher backgrounds and the long RPC
dead times. This requires development of a new detection technique,
which should be robust, inexpensive and capable of coping with high
backgrounds.

In this paper, we present a scintillator-based solution for the Belle
II endcap KLM (EKLM) detector and demonstrate that it matches the
scientific and environmental requirements. The prospects for this
technology were first discussed in a previous paper~\cite{ITEP_Strip},
after which this technology was chosen as the baseline technology for
Belle II. Considerable R\&D studies have been performed in order to
demonstrate the feasibility of the proposed approach, and the
construction of the new EKLM detector is already underway. A similar
approach was also proposed for the SuperB experiment in
Italy~\cite{superb}, which has since, quite unfortunately, been
terminated.

\section{The Belle II EKLM system}

The Belle II EKLM system will follow its Belle predecessor, consisting
of alternating layers of active charged particle detectors and 4.7~cm
thick iron plates. The iron plates serve as the magnetic flux return
for the solenoid and provide a total of 3.9 interaction lengths of
material for a particle travelling normal to the detector
planes. There are 14 detector and 15 iron layers in each of the
forward and backward endcaps. Each detector layer is divided into 4
sectors, which can be separately installed. Since the Belle iron
structure will be retained, we plan to simply swap in our new
detector, subject to the constraint that it must mate with the
existing sector frames which guide installation into the 4.0~cm wide
iron gaps.

Scintillator counters with wave-length-shifting (WLS) fibers as a
readout option for the EKLM were first proposed by our group in the
Belle upgrade Letter of Intent~\cite{LoI}. Charged particle detection
with such detectors, equipped with photomultiplier tube (PMT) readout,
is a well-established technique~\cite{Minos, Opera}. However, in the
case of the Belle II detector, the limited space and strong magnetic
field do not allow use of PMTs. As an alternative photodetector, we
proposed multipixel silicon photodiodes operating in the Geiger mode,
originally developed in Russia~\cite{SiPM} in the 1990s, and currently
produced by many companies under different names: Silicon
PhotoMultiplier (SiPM), Avalanche Photo-Diodes (APD),
Metal-Resistor-Semiconductor (MRS), Multi-Pixel Photon Counters
(MPPC), Multi-Pixel Avalanche Photo-Diodes (MAPD), among others. In
this paper, we will use the generic name SiPM to include all such
detectors.

SiPMs allow for compact detectors and uncompromised operation in
strong magnetic fields. The first large-scale (7620 channels) SiPM
application was the CALICE experiment's hadron
calorimeter~\cite{Danilov, Sefkow}, which demonstrated the feasibility
to use SiPMs experimentally, as well as their advantages over
traditional PMTs. The use of SiPMs in a real experiment with a huge
number of readout channels ($\sim$65~k) began with the near detector
of the T2K experiment~\cite{T2K}, where many subsystems are based on
SiPM readout of scintillator light. However, the background rates and
the radiation environment in neutrino experiments are much more
benign, and therefore more stringent testing is required to prove the
applicability of this technique to the more exacting Belle II
environment.

\subsection{General layout}

The base element of the new detector system is a scintillator strip of
polystyrene doped with a scintillator die. It has a rectangular
cross-section and a varying length of up to 2.8~meters to match the
geometry of an individual sector. The strip height is limited to 10~mm
by the mechanical constraints of the gap between the iron yoke and
frame structure. The selected strip width of 40~mm is a compromise
between the desire for a moderate number of channels and the required
spatial resolution for muon and \kl\ reconstruction. The granularity
is similar to the average granularity of the Belle experiment's
original RPCs. It is commensurate with the uncertainties due to muon
multiple scattering and typical hadron shower transverse sizes;
further increase of the granularity does not improve muon
identification performance and \kl\ angular resolution. The hit
registration efficiency does not depend on the strip width as long as
the rates remain below data acquisition (DAQ) and SiPM saturation
levels.

\begin{figure}[htb]
\begin{center}
\includegraphics[width=0.92\textwidth]{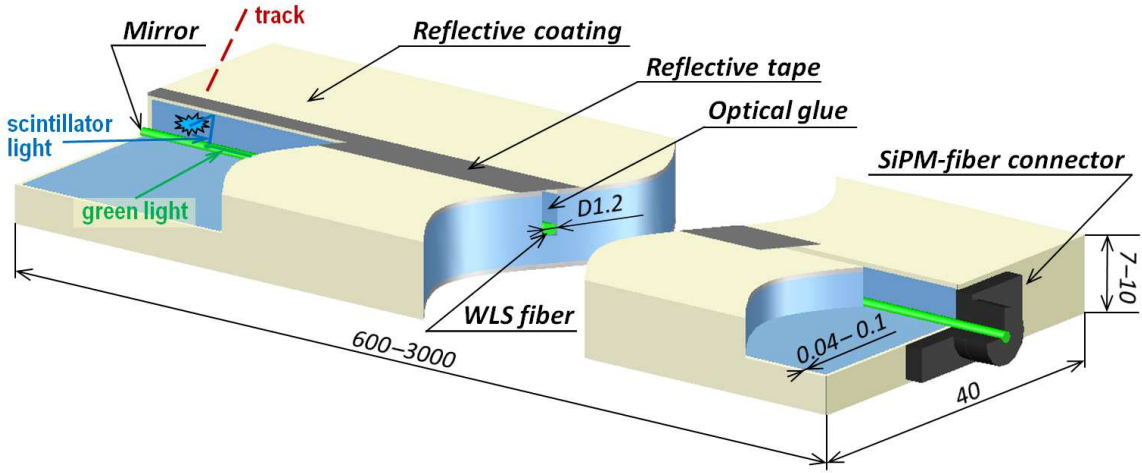}
\end{center}
\caption{Schematic view of the scintillator strip. Dimensions are in
  mm.}
\label{fig:strip}
\end{figure}

A schematic of the assembled strip is shown in Fig.~\ref{fig:strip}.
Individual strips are covered with a diffuse reflective coating; each
strip has a groove in the center to accommodate a WLS fiber.
Scintillator light is collected by a WLS fiber and transported to the
photodetector. Each WLS fiber is read out from one (near) side -- the
far end of the WLS fiber is mirrored to increase the total light yield
from the strip. To increase the efficiency for light collection, the
WLS fiber is glued to the scintillator with optical glue. The SiPM is
then coupled to the fiber end, and fixed and aligned with the fiber
using a plastic housing.

\begin{figure}[htb]
\begin{center}
\includegraphics[width=0.92\textwidth]{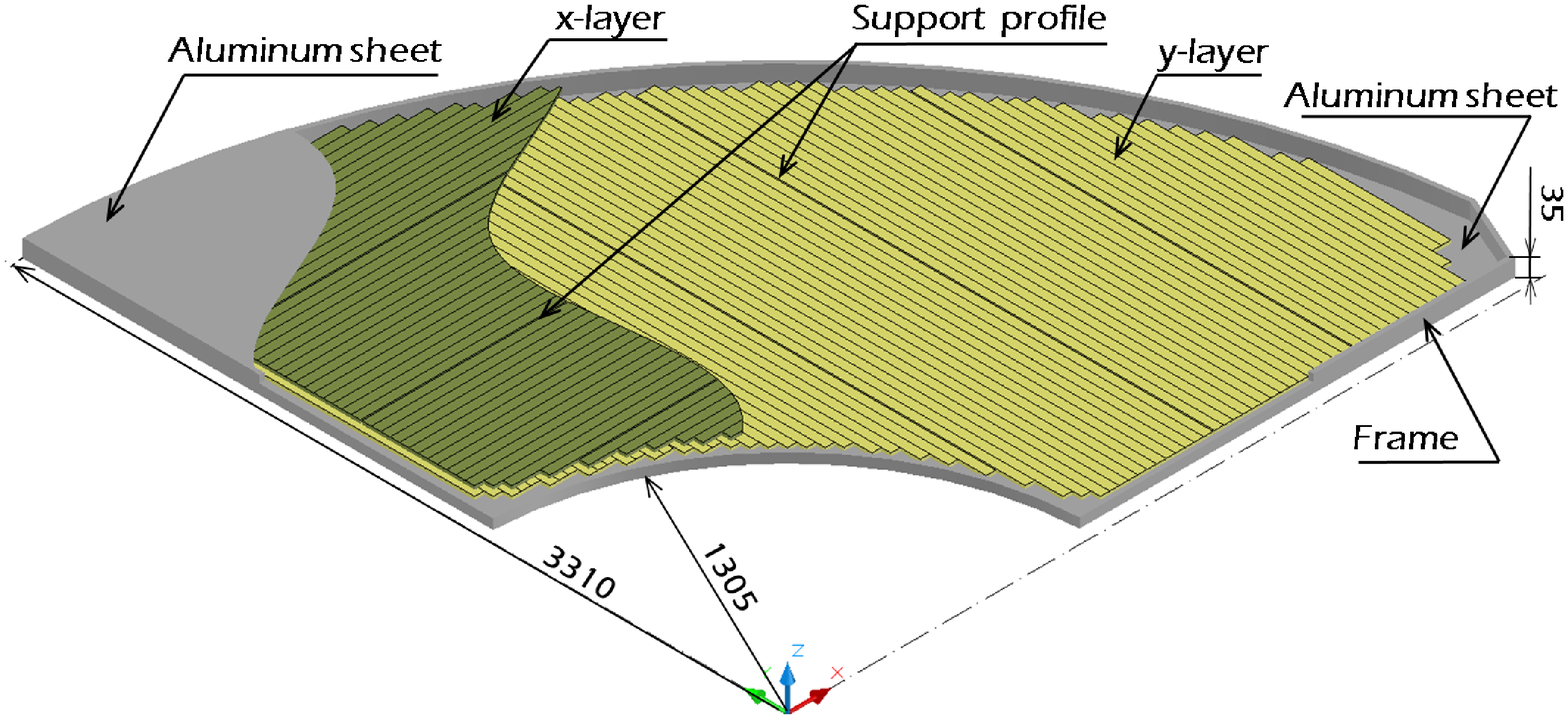}
\end{center}
\caption{Schematic view of one superlayer formed by scintillator
  strips. Sizes are given in mm.}
\label{fig:sector}
\end{figure}

One superlayer is formed from two fully overlapping orthogonal layers,
each containing 75 scintillator strips. The independent operation of
two planes in one superlayer should reduce the combinatorial
background in comparison with the present RPC design, where every
background hit produces signals in both readout planes. Scintillator
strips are arranged in a sector, with a geometry matched to the
existing gap in the iron yoke, as shown in Fig.~\ref{fig:sector}.
Fifteen strips are glued to a common thin (1.5~mm) polystyrene
substrate from both sides and fixed in the support profile. In
addition to providing rigidity and moderate passive neutron shielding,
this substrate also serves to absorb protons scattered by background
neutrons, and thereby prevents correlated hits in two layers appearing
from a single background neutron.  The sector frame is the same as
that used for Belle's RPC mounting. The dead zone around the inner arc
is estimated to be $\sim$1\% of the total sector area due to the
inscription of the rectangular structure into the circular housing;
this inner dead zone is approximately the same as that of the Belle
RPC EKLM detector. Around the outer circumference, the dead zone is
$4\%$, primarily owing to the presence of front-end electronics and
cables. In practice, this dead area cannot be recovered, as the
extremely short strips that might be installed there have very small
coverage, and do not justify the additional number of read-out
channels they would entail.\footnote{We note that the acceptance loss
  at large radii is not critical for muon and \kl\ reconstruction.} In
the middle part of the sector, unavoidable small dead zones ($0.8\%$)
are due to the presence of support structures. The total insensitive
area between strips due to the reflective cover is only $0.3\%$. In
total, the geometrical acceptance of the new system is slightly better
than that of the Belle RPC EKLM.

In the new EKLM system the scintillator-based superlayers are
installed in all 14 gaps in the magnet yoke in both the forward and
backward endcaps. The entire system consists of 16800 scintillator
strips of varying lengths.

\subsection{SiPMs, detailed description}

\begin{figure}[htb]
\begin{center}
\includegraphics[width=0.9\textwidth]{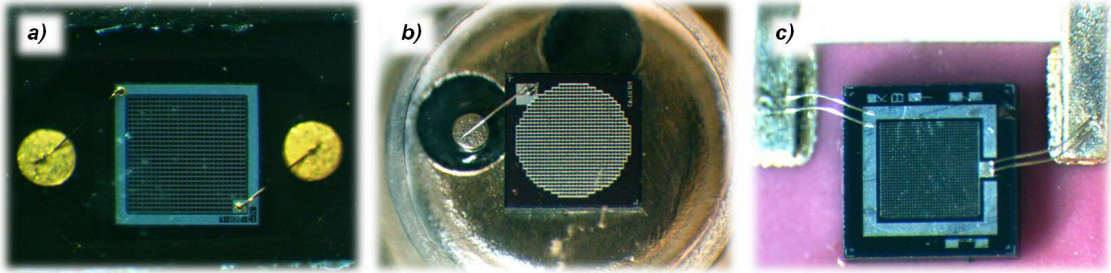}
\end{center}
\caption{Microscopic views of SiPM surfaces from different
  manufacturers: a) Hamamatsu (MPPC S10362-13-050, $1.3\times
  1.3$~mm$^2$)~\cite{hamamatsu}, b) CPTA (CPTA 143,
  $d=1.3$~mm)~\cite{cpta} , c) MEPhI/PULSAR ($1.0\times
  1.0$~mm$^2$)~\cite{mephi}.}
\label{fig:photo}
\end{figure}

A SiPM is a matrix of tiny photo-diodes (pixels) connected to a common
bus and operating in the Geiger mode~\cite{SiPM}. A photograph of the
SiPM's surface, comparing samples from different vendors, is shown in
Fig.~\ref{fig:photo}.
The number of pixels is $\sim$$100 - 1000$, over a typical area of $1
\times 1$~mm$^2$. Connection to the common bus is generally achieved
via a $\sim$M$\Omega$ resistor. This resistor limits the Geiger
discharge developed in a pixel by photo- or thermal electrons. In this
mode, the SiPM response depends on the number of fired pixels and is
proportional to the initial light, as long as the number of fired
pixels is much smaller than the total number of pixels.

As compared with conventional vacuum photomultipliers, SiPMs have
lower operating voltages, typically of the order a few tens of volts,
which is higher than the breakdown threshold by several volts. The
values of the overvoltage and the capacitance of a single pixel
determine the photo-detector gain (typically of order $10^6$). The
values of quenching resistance and the capacitance of a single pixel
determine the dead time of a pixel, which is typically
$\sim$$10-100$~ns. The photon detection efficiency ($PDE$) depends on
the overvoltage and reaches $30-40\%$ for modern devices. SiPMs with a
smaller number of pixels usually have higher $PDE$ values owing to the
smaller non-sensitve area between pixels. The insensitivity of our
SiPM samples to magnetic fields was checked in fields up to
4~T~\cite{Andreev}.

Among the disadvantages of SiPMs are the high levels of noise
($\sim$$10^5-10^6$~Hz/mm$^2$ at a 0.5 photoelectrons (p.e.)
threshold), an optical inter-pixel cross-talk producing a tail of
large noise amplitudes, and high sensitivity of the SiPM response to
ambient temperature, since the breakdown voltage depends on
temperature ($\sim$60~mV/K for Hamamatsu, $\sim$20~mV/K for CPTA).

A comparison of the measured SiPM characteristics from different
vendors is shown in Fig.~4.
\begin{figure}[htb]
\begin{center}
\includegraphics[width=0.92\textwidth]{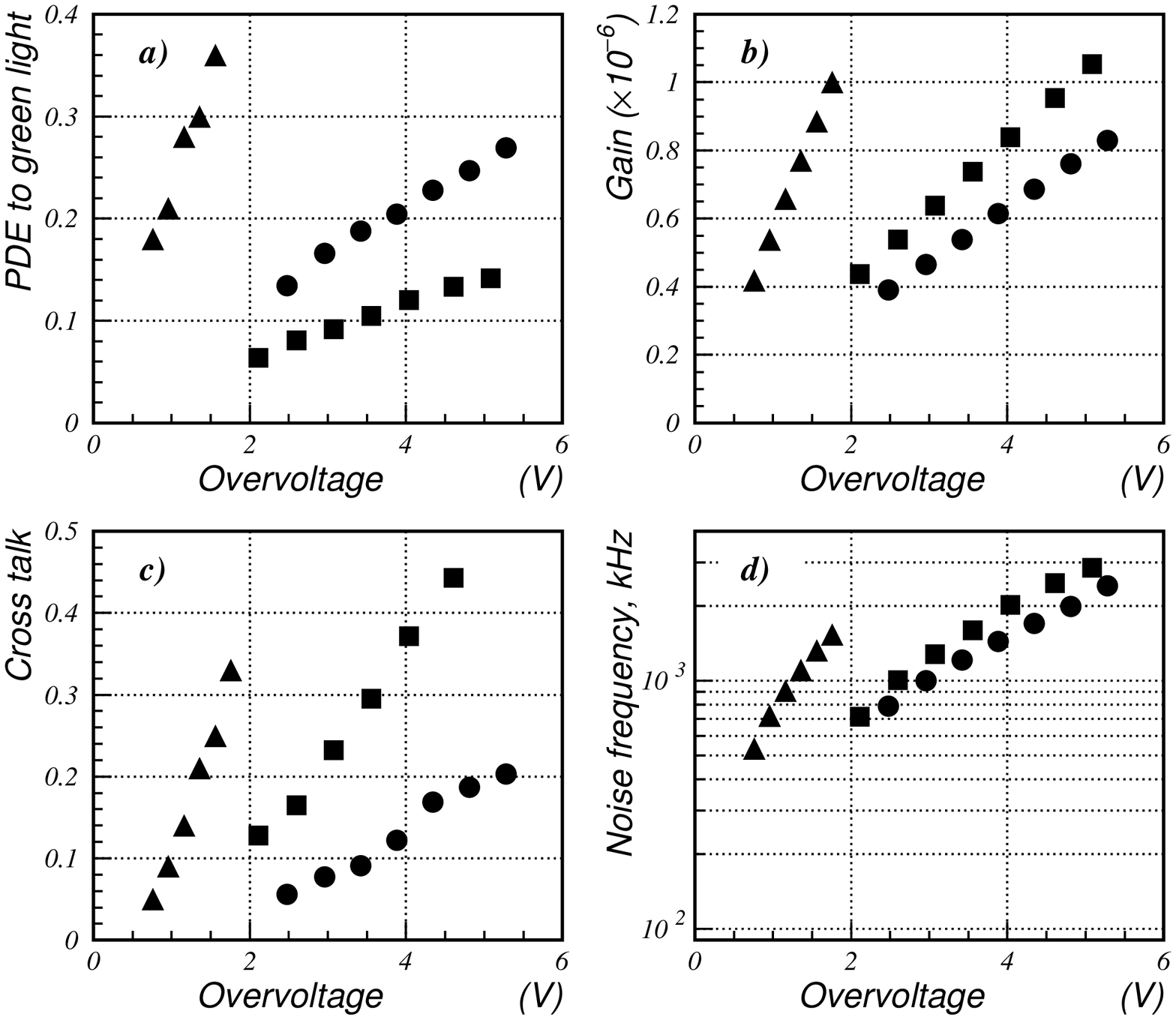}
\caption{Overvoltage dependence of SiPM a) efficiency, b) gain, c)
  cross talk and d) noise frequency for photo-detectors from different
  manufacturers: Hamamatsu (triangles), CPTA (circles), MEPhI/PULSAR
  (squares).}
\end{center}
\label{fig:Sipm}
\end{figure}
The photon detection efficiency for green light from the Y11 WLS
fiber, gain, cross-talk, and noise rate are shown as a function of
overvoltage for detectors produced by MEPhI/PULSAR (Russia), CPTA
(Russia) (CPTA 143), and Hamamatsu (Japan) (MPPC S10362-13-050). For these
measurements, we used a calibrated light source which was being
monitored with a standard PMT. The presented characteristics are
extracted from fits to the randomly triggered (to collect noise) and
externally triggered (synchronized with light source pulses) ADC
spectra. The spectrum of the number of fired pixels for externally
triggered data is fitted to a Poisson distribution convolved with a
cross-talk correction. In this way, we extract both $PDE$ from the
Poisson mean, and the cross-talk value independently. To minimize
statistical fluctuations, the data shown are averaged over 10-50
specimens from each vendor. While the breakdown voltage varies for
individual specimens, the dependence on overvoltage ($\Delta V = V -
\hvb$), in general, has a very small dispersion; \hvb\ itself can be
determined from extrapolation assuming a linear gain dependence on
voltage.

MEPhI/PULSAR SiPMs were developed much earlier than other mass
produced SiPMs, and their efficiency is correspondingly somewhat lower
than the more advanced models. SiPMs produced by CPTA and Hamamatsu
meet our minimum requirements, defined as high detection efficiency
for Minimum Ionizing Particles (MIP) and low cross-talk. In spite of
the higher observed single pixel noise for CPTA SiPMs, their smaller
cross-talk value results in a relatively low noise rate, of the order
of kHz, at a 7.5~p.e. threshold. It should be noted that the noise
rate is not critical for our system, since at the chosen threshold,
the main background arises from neutrons, rather than SiPM noise. Both
CPTA and Hamamatsu were considered as the primary SiPM vendors for the
T2K experiment, and both vendors have produced thousands of SiPMs that
mate to our 1.2~mm diameter fiber. Our choice between these two (both
acceptable) vendors was made based on a radiation hardness study as
described in Section~\ref{sipm_rad}.

\subsection{Scintillator strips}

The largest light yield in the blue regime is achieved with organic
scintillators, {\it e.g.} those produced by Bicron~\cite{bicron}.
However, the production technique for such scintillators does not
allow production of individual long strips, which can only be achieved
by gluing together short strips, and is also very expensive to be
practically considered for a detector having the large area of the
Belle II EKLM system ($>1000$~m$^2$). Therefore, we have focused on
achieving sufficient light yield with much less expensive polystyrene
scintillators.

Scintillator strips made of polystyrene, typically doped with PTP
(p-terphenil) or PPO (2,5-diphenyloxazole) and POPOP
(1,4-bis-2-(5-phenyloxazolyl)-benzene), can be produced by extrusion,
in principle allowing manufacture of very long strips. Although
different vendors use slightly different ingredients and production
techniques, they generally achieve quite similar quality, and have
similar prices. We tested the characteristics of scintillator strips
from three vendors: Amkris-Plast (Kharkov, Ukraine)~\cite{kharkov},
which produced the strips in use for the Opera
experiment~\cite{Opera}, Fermilab (USA)~\cite{fermilab}, and also
Uniplast (Vladimir, Russia)~\cite{vladimir}, which both produced the
scintillator strips for the T2K experiment~\cite{T2K}. All strips had
width 40~mm. The height of the Amkris-Plast and Fermilab strips was
10~mm, and the thickness of the co-extruded TiO$_2$+polystyrene
coating was $\sim$100~microns. For the third vendor, we used existing
Uniplast strips produced for the T2K experiment with 7~mm
thickness. The reflection coating is achieved in this case by chemical
etching of the surface and is thinner ($40-50$~$\mu$).

In all tested strips, Kuraray WLS Y-11(200)MSJ multi-cladding $1.2$~mm
diameter fibers~\cite{Kuraray} were glued using SL-1 glue produced at
SUREL (St.~Petersburg, Russia)~\cite{surel}. The Kuraray fiber
provides more light output than other fibers and also has a favorably
long attenuation length. This fiber has been used in many detectors
with geometry and scintillator plastic similar to ours~\cite{Minos,
  Opera, T2K}.  The light emission spectrum of the Kuraray Y11 fiber
is, in fact, quite well-matched to the SiPM spectral efficiency.

The strips have been tested using a cosmic ray stand. The cosmic ray
trigger is provided by a pair of short trigger strips ($L=16$~cm)
placed above and below the strip being measured. The trigger strips
were moved along the tested strip to measure the light yield as a
function of the distance to the photodetector. All strips were tested
using SiPMs from the same vendor in order to minimize systematic
errors. To determine the average number of detected photons, we
correct the measured number of fired pixels by the cross-talk fraction
($1/(1+\delta)\!\sim\!0.85$) and conservatively use a truncated mean
of the Landau distribution, discarding 10\% of the lower portion of
the distribution and 30\% of the higher portion, then take the average
of the resulting samples. The resulting averaged number, which is very
close to the maximum of the spectra, is then corrected by a factor of
0.87 for oblique incidence of cosmic-ray muons, determined using a toy
Monte Carlo simulation.

\begin{figure}[htb]
\begin{center}
\includegraphics[width=0.92\textwidth]{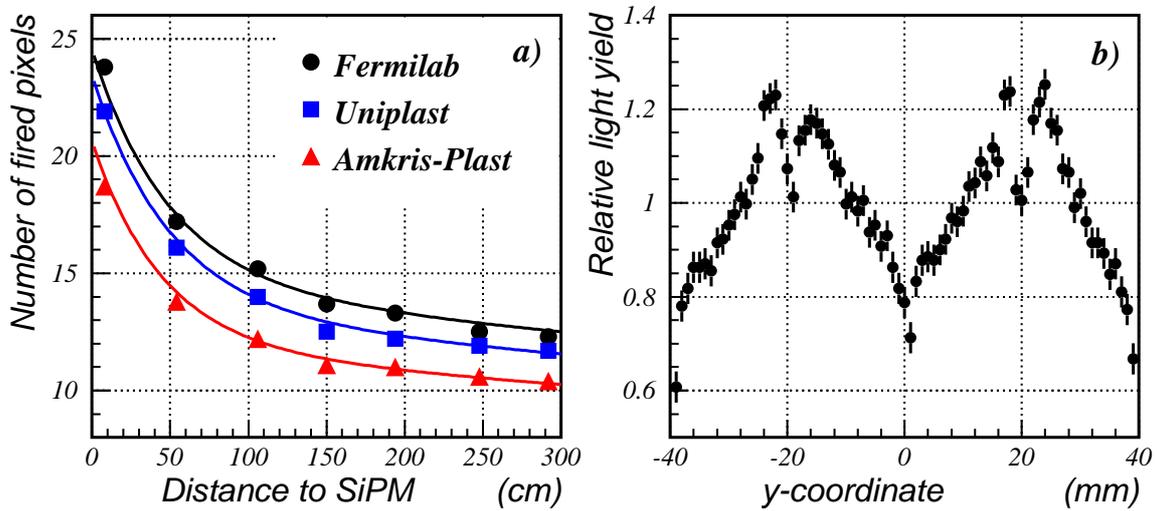}
\end{center}
\caption{a) Distribution of the average number of SiPM-detected
  photons as a function of the distance to the photodetector. b) The
  dependence of the relative light yield on the transverse position of
  the hit for two neighboring strips.}
\label{fig:dist}
\end{figure}

The distribution of the average number of detected photons is shown in
Fig.~\ref{fig:dist}a as a function of the distance to the
photodetector (Hamamtsu MPPC S10362-13-050).
We compare our results with the performance of the Fine-Grained
Detectors (FGD) of the T2K experiment~\cite{T2K_tdr}, where strips of
the same length and the same SiPMs were used similarly. The
scintillator light at the FGD is collected from the Fermilab strips
($10 \times 10$~mm$^2$) using diameter 1~mm fibers. The light yield
achieved is presented in the FGD TDR (Fig.~6 of Ref.~\cite{T2K_tdr}).
Those tests were done in 120~MeV/$c$ beams and then corrected, in the
case of both muons and pions, for the expected MIP response. The light
yield dependence on the distance to the photodetector in our case
nicely agrees with the FGD test results. The light yield at the near
end of the strip is $\sim$21 photoelectrons/MIP for FGD detector. We
achieve a slightly higher light yield in spite of the four times
larger strip width in our case (the light yield decreases with
increasing width). We attribute this significant improvement in the
light collection efficiency to the larger diameter of the fiber (light
yield $\propto$ diameter) and the gluing of the fiber to the strip.

The Fermilab and Uniplast strips, which have the same price per unit
weight, provide almost the same light yield at the SiPM, despite the
fact that the Uniplast strips are 30\% thinner. Apart from
considerations of cost and weight of the entire system, the smaller
scintillator thickness is beneficial in terms of reducing the neutron
background. Indeed, the neutron rate is proportional to the
scintillator volume. This consideration motivated our final selection
of the Uniplast strips for the Belle II EKLM construction.

The transverse uniformity of the Uniplast strip response was measured
using tightly collimated and triggered $^{90}$Sr radioactive
sources. Figure~\ref{fig:dist}b shows the relative pedestal-subtracted
ADC response, which is proportional to the light yield, for two strips
mechanically pressed against each other. The nonuniformity of the
response is $\pm 25\%$. A reduction in the number of registered
photons is observed near the groove, where the effective thickness of
the scintillator is smaller, and at the strip edge, where the
efficiency for scintillator light to reach the WLS fiber is
smaller. From these results, we conclude that the light yield is
adequate, irrespective of where a particle hits the strip.

In the same test we also measured the optical cross-talk between two
neighbouring strips. We used the ADC spectrum of the non-irradiated
strip, when the collimated beam was placed at least 5~mm away from the
common edge (otherwise the angular divergence of the beam may result
in real scintillation in the strip under consideration). From the
measured yield, we conclude that the optical cross-talk is $(3 \pm
1)\%$.

We measure the time resolution using a cosmic-ray trigger for a strip
read out by two SiPMs, one on each fiber end. A TDC was started by one
SiPM signal and then stopped by the second. From the observed time
difference distribution, we can derive a time resolution $\sigma_t
\approx 0.7$~ns. The measured velocity of light propagation in the WLS
Y11 Kuraray fiber with 1.2~mm diameter is $v=17$~cm/ns, which
translates into $\sim$12~cm of spatial resolution along the strip.

\section{Long-term stability and radiation hardness}

Scintillator counters with WLS fiber using standard PMT read-out are
used in many existing detectors, including those at hadron colliders,
where the radiation dose is a few orders of magnitude higher than
those anticipated at SuperKEKB. Such counters typically operate
without degradation for decades. For example, the light collection of
electromagnetic calorimeters for the HERA-B and LHCb experiments is
based on Uniplast scintillator tiles read out by Kuraray Y11
fibers~\cite{Hera-b}. During several years of operation, no
deterioration of the electromagnetic calorimeter performance was
observed~\cite{hardness}.

SiPMs demonstrate excellent long-term stability. Our experience in
operating the 7620-channel CALICE hadron calorimeter prototype using
scintillator/WLS fiber/SiPM detecting tiles during three-year beam
tests at CERN and FNAL~\cite{Danilov, Sefkow} showed no significant
deterioration of performance. Only 8 dead channels ($0.1\%$) were
observed, while the characteristics of the remaining channels were
acceptably stable. The long-term stability of Hamamatsu SiPMs was
tested in the T2K experiment, where $\sim$65 thousand channels were
studied during a few years of operation. Failure was observed in
$\sim$20 channels, {\it i.e.} less than $0.03\%$, primarily owing to
mechanical damage.

However, the adequacy of SiPMs radiation hardness for our application
needs to be proven experimentally. In particular, the radiation
hardness of SiPMs was measured to be high in the case of irradiation
with electrons and photons, while neutrons and protons cause
significant damage at a moderate integrated dose of
$\sim$1~kRad~\cite{Danilov, Tarkovsky}. Typical radiation damage
consists of defect formation in the thin area from which charge
carriers are collected, resulting in an increase of the SiPM noise
rate and, consequently, the dark current. Overall, the SiPM dark
current grows linearly with particle flux as in other silicon
detectors.

\subsection{Study of scintillator strips aging}

The long-term stability of the light collection efficiency of strips
with glued WLS fibers was determined using a test module consisting of
96 strips (1 meter long), manufactured in 2006. This test module was
used for the measurement of the neutron background in the KEKB tunnel
for a period of 6 months. Modules were placed in a `hot' region and
accumulated a dose corresponding roughly to 3-5 years of Belle II
operation. In 2009, these strips were retrieved and examined for
radiation damage and aging. No measurable degradation of the light
yields (using new, not irradiated SiPMs) within the 10\% accuracy of
the measurements was observed three years after production. We also
subjected a short strip with glued WLS fibers to fast aging in a
thermostat at a temperature $80^\circ$~C during 4 months
(corresponding to 10 years at room temperature according to the Van
tHoff's rule) and again found no noticeable aging effects.

\subsection{SiPM's aging}
\label{sipm_rad}

Given the relative novelty of SiPMs, we have studied their radiation
hardness in detail, including any possible deterioration as a function
of particle type to which they are subjected.  Radiation damage from
protons and neutrons is energy-dependent and difficult to predict
without knowledge of the beam background spectra. Therefore, to
measure possible damage to SiPMs due to the high radiation levels
anticipated at SuperKEKB, we performed radiation tests directly in the
KEKB tunnel. Eight SiPMs produced by Hamamatsu and eight CPTA SiPMs
were placed, for 6 weeks, in the KEKB tunnel at a location where the
dose is much higher than in the nominal SiPM position. The integrated
neutron dose at the tunnel measured with Luxel badges
(J-type)~\cite{luxel} was $\sim$1.2~Sv. After irradiation, the SiPM
dark current increased for both Hamamatsu and CPTA devices by a factor
of 3.6 with small dispersion ($\sim$15\%), as measured in 8 specimens.

The neutron dose was also measured near the nominal {\it in situ}
position of the SiPM. The measured dose varies from 0.8 to
2.0~mSv/week at a luminosity $\mathcal{L} \! \sim \! 2 \times
10^{34}$~cm$^{-2}$s$^{-1}$. Assuming conservatively that the neutron
dose increases linearly with luminosity, the expected neutron dose
integrated over 10 years of SuperKEKB operation at $\mathcal{L} \!
\sim \! 8 \times 10^{35}$~cm$^{-2}$s$^{-1}$ does not exceed
40~Sv\,\footnote{Monte Carlo simulation of the radiation dose actually
  predicts a dose which varies more slowly than linearly.}. The
extrapolated factor of dark current increase after 10 years of
SuperKEKB is therefore $\sim$120. In the case of the Hamamatsu SiPMs,
the expected dark current is $\sim$12~$\mu$A, In the case of the CPTA
SiPMs, the dark current becomes much higher ($\sim$170~$\mu$A), since
the initial noise is higher for the CPTA SiPMs. The difference between
the Hamamatsu and CPTA SiPMs can be explained by the different
thicknesses of the sensitive zone and the difference in manufacturing
details (purity of the raw materials and also of the SiPM surface) on
which both the initial noise and the performance depend.

The study of SiPM properties after irradiation was done at the
Institute of Theoretical and Experimental Physics (ITEP) in
Moscow. The SiPMs were subjected to fast irradiation in the
200\mev\ secondary proton beam of the ITEP synchrotron
($\sim$$10^{10}~p/$cm$^2/$hour). One month after irradiation, the
Hamamatsu SiPM dark current was measured to be 16~$\mu$A
(corresponding to more than 10 years of operation at SuperKEKB) for
two specimens subjected to an integrated proton flux $2.0 \times
10^{11}~p/$cm$^2$ and 6.5~$\mu$A (corresponding to 5 years of
operation at SuperKEKB) for two specimens subjected to an integrated
proton flux $7.4 \times 10^{10} ~p/$cm$^2$.

We studied the light yield from minimum-ionizing particles traversing
the far end of the strip ($\sim$3~m from the photodetector) detected
with both non-irradiated and also irradiated SiPMs. The average number
of photoelectrons (p.e.) obtained from a MIP signal using irradiated and
control SiPMs is unchanged, though in the case of irradiated SiPMs,
the MIP signal is slightly smeared owing to increased noise. The
dependence on the threshold of the efficiency at the far end of the strip,
as well as the noise rate, in terms of the number of photoelectrons, is
presented in Fig.~6. We conclude that the light detection efficiency
is not significantly changed after irradiation, while the internal
noise rate increases significantly. However, at a threshold of
$\sim$7.5 fired pixels the background rates from increased SiPM
noise remains smaller ($\lesssim$10~kHz) than the physical background
rate due to neutron hits in the scintillator ($\lesssim$160~kHz for
the longest strip). The MIP detection efficiency is similar with
irradiated CPTA's SiPM, but the noise rate becomes unacceptably high
($100-500$ kHz).
\begin{figure}[htb]
\begin{center}
\includegraphics[width=0.92\textwidth]{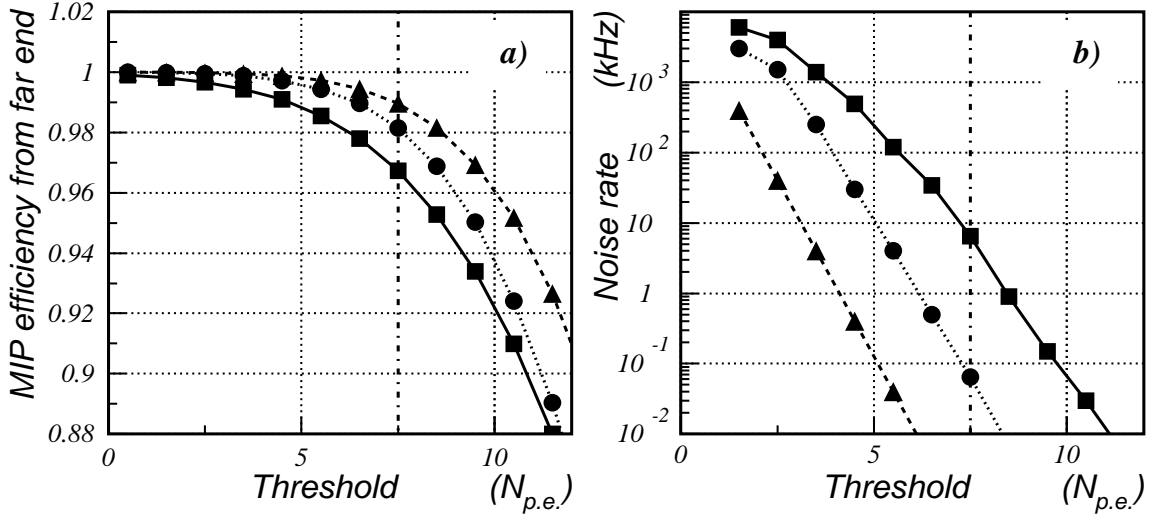}
\caption{Dependence on the threshold (defined by the number of
  photoelectrons) of a) the MIP detection efficiency from the strip
  far end and b) the internal noise rate for Hamamatsu SiPMs. The
  triangles correspond to the non-irradiated SiPMs; circles correspond
  to SiPMs irradiated with a $7.4 \times 10^{10}~p/$cm$^2$ dose
  (corresponding to 5 years of operation at SuperKEKB) and the squares
  show the case for an SiPM irradiated with $2.0 \times 10^{11}
  ~p/$cm$^2$ dose (more than 10 years of operation at SuperKEKB). }
\end{center}
\label{fig:Rad}
\end{figure}

We therefore conclude that the Hamamatsu's SiPM radiation hardness is
sufficient for successful EKLM operation at the design SuperKEKB
luminosity for at least 10 years. The significantly increased dark
current, however, should be considered in the design of electronics,
HV supplies, slow control and the calibration procedure. This work is
currently underway.

\section{Improving light collection efficiency}

In the previous Sections, we have demonstrated that commercially
available SiPMs provide the basis for an adequate KLM detector system
at the Super KEKB factory. However, given the expected dispersion of
strip characteristics during mass production, and to also ensure
better robustness and physics performance, we further studied the
possibility of improving the light collection efficiency at the
SiPM. We carefully studied two points in the system where there may be
loss of light: in the transport of scintillator light to the WLS
fiber, and the subsequent conveyance of the WLS fiber light to the
SiPM.

The effective transport of the scintillator light to the WLS fiber is
improved with the help of the optical glue, which has refractive index
similar to that of both the strip and the outer cover of the
fiber. The glue significantly reduces the light loss due to the
roughness of the mating surface.  However, it is very difficult to
fill the corners of the rectangular groove even with very
low-viscosity glue, resulting in small residual air bubbles at the
groove corners, and therefore enhanced scattering and absorption of
photons in this region. This problem does not arise if the shape of
the groove is rounded. By milling the groove to a rounded shape, our
lab tests demonstrate an increase in the average detected light yield
by $(25\pm 5)$\% relative to the ``standard'' rectangular-shaped
grooves.

\begin{figure}[htb]
\begin{center}
\begin{tabular}{cc}
\includegraphics[width=0.47\textwidth]{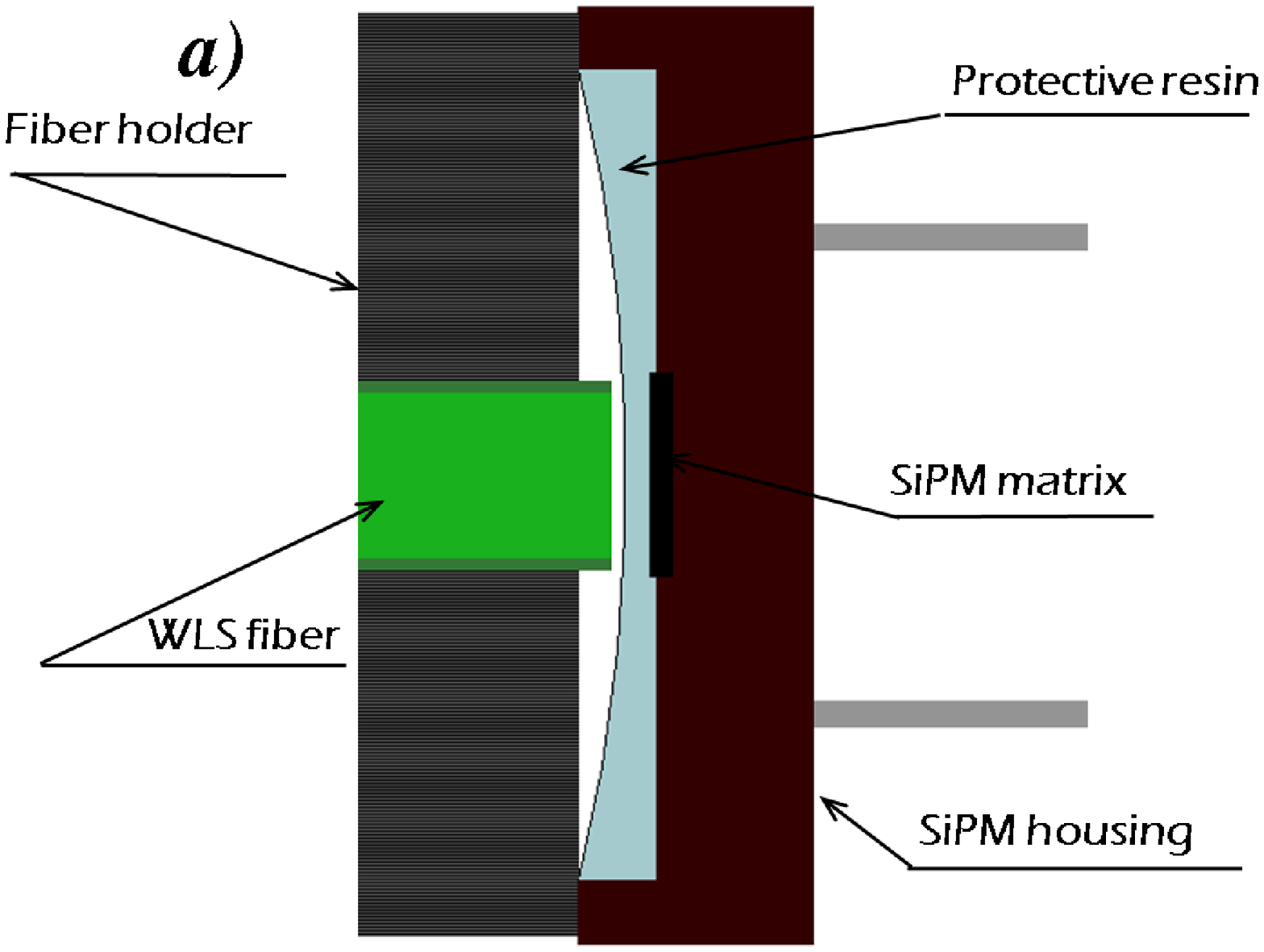}  &
\includegraphics[width=0.46\textwidth]{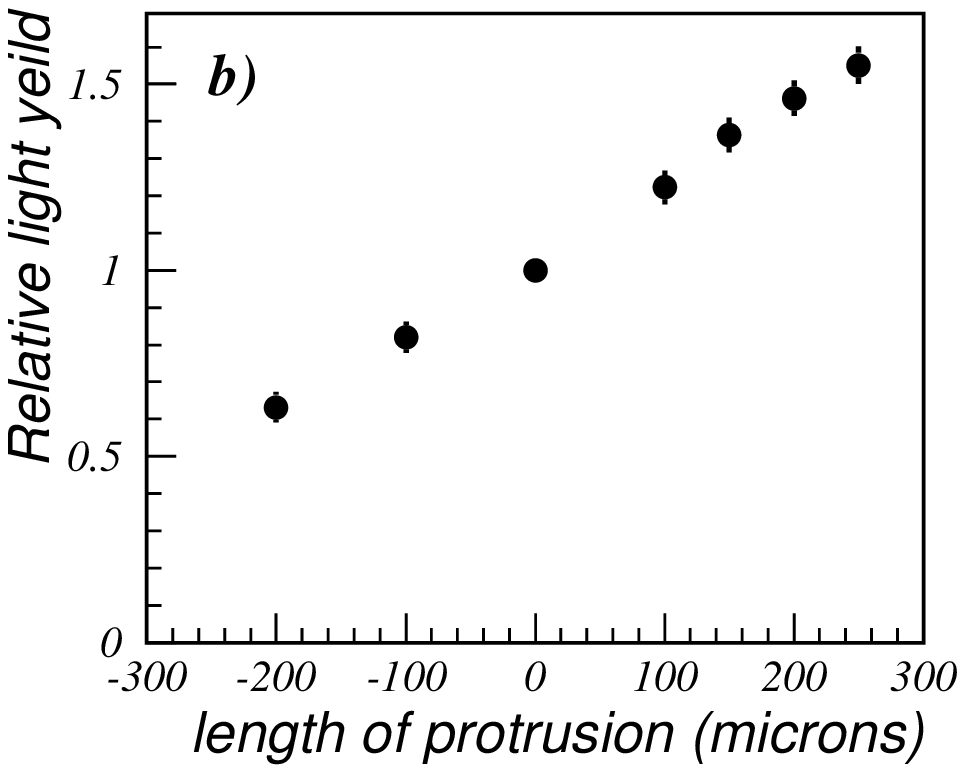}
\end{tabular}
\end{center}
\caption{ a) schematic view of the optical connection of the WLS fiber
  end and SiPM (Hamamatsu MPPC S10362-13-050); b) relative light yield
  dependence on the length of the protrusion (normalized to zero
  protrusion).}
\label{fig:surf}
\end{figure}

The optical contact between the WLS fiber end and the SiPM can be also
improved. The SiPM matrix is covered with a thin protective resin.
The optimal configuration of the optical contact corresponds to the
minimal air gap between the fiber end and this cover, while avoiding
mechanical contact so as not to damage the cover. We studied the
flatness and thickness variation of this cover using a microscope, and
found that the shape of the cover in the case of the Hamamatsu SiPMs
is not flat, but concave (see Fig.~\ref{fig:surf}a), which is
consistent with surface tension effects during the hardening of the
resin. As a result, the light from the fiber is defocused at the SiPM
matrix -- this effect is actually aggravated in practice, as the air
gap turns out to be much larger than that expected for the flat
cover. We measured the dependence of the light yield at the SiPM on
the length of the fiber inside the SiPM (Fig.~\ref{fig:surf}b). This
effect turns out to be substantial: a 150 micron extension to
compensate for the concavity of the resin (which still ensures no
mechanical contact of the fiber end to the resin surface) results in a
$(37 \pm 5)$\% increase of the number of photoelectrons detected at
the SiPM.

As a result of these studies, we improved the light collection
efficiency by $\sim$70\%, by implementing the straightforward
modifications to the strip production technology described above.

\section{Physics Performance}

Besides the main mission to detect \kl\ mesons and muons, another
important task for the KLM system at Belle II, motivated by our
extensive Belle experience, is to serve as an hadronic veto to improve
the signal to background ratio for the so-called missing energy modes,
such as $B^+ \to \tau^+ \nu$, $B \to K^{(*)} \nu \bar{\nu}$,
etc~\cite{belleii-physics}. Indeed, the major background remaining
after baseline selection for these modes is due to $B$-decays with
missing neutrals or un-reconstructed charged hadrons. In this case, a
veto on KLM hadronic clusters would be very useful, provided a high
efficiency and low background can be achieved with the new KLM system.

The muon identification and hadron cluster reconstruction performance
depends on the background rate, mostly related to slow neutrons
produced in the accelerator tunnel and in the vicinity of the
interaction region. In spite of the much higher sensitivity to
neutrons for scintillators containing hydrogen compared to the
gas+glass based RPC, the total background rate of the new EKLM system
is expected to be much lower than those achievable with RPC
design. Indeed, single background neutrons are usually absorbed in one
scintillator strip, so there are no two-dimensional hits from single
background particles; fake two-dimensional hits mostly arise from
random coincidences within the integration time. This is in contrast
to RPCs, where even a single scattered neutron induces signals on both
stereo strips.

To estimate the background rate we have used two independent methods:
i) a conservative extrapolation of the measured neutron rates using a
scintillator test module, and ii) a Monte Carlo detector study.

\subsection{Measurement of the background neutron rate}
\label{sect:bg}

As RPC and scintillators have different responses to beam backgrounds,
an extrapolation of the known Belle background rate at the RPC to the
expected rate at the scintillator KLM at SuperKEKB is difficult. We
have therefore performed measurements of the neutron flux using a
scintillator test module installed directly in the KEKB tunnel close
to the EKLM. The test module contained four layers. Each layer,
surrounded by a copper shielding box (4 cm thickness) for protection
from external light and soft electron/photon backgrounds, was
assembled from 24 strips. Each scintillator strip, with a WLS fiber
and SiPM readout, had a size of $1000 \times 40 \times
10$~mm$^3$. During KEKB operation, the signals from 96 channels were
collected by random triggers using CAMAC ADCs, with a gate set to be
100~ns.

We found that a simple shielding box was inadequate for protecting the
module from numerous charged tracks and showers. To discriminate
neutron signals from other sources, different hit patterns in the four
layers were used. A significant portion of the observed events (where
at least one SiPM signal exceeds the threshold of 0.5 MIP) had
multiple hits from charged tracks or showers; to suppress such
backgrounds, we required a veto in the two outer layers and also
required a single hit in one of the two inner layers. The derived
neutron rate at a nominal 0.5 MIP threshold, at a luminosity of
$\mathcal{L}\! \sim\!1.4 \times 10^{34}$~cm$^{-2}$s$^{-1}$, was
6~Hz/cm$^2$. We use this number to estimate the expected occupancies
in the readout electronics and trigger (maximum 160~kHz from the
longest strip, 60~kHz averaged), as well as to estimate the expected
backgrounds for \kl\ reconstruction.

\subsection{Monte Carlo simulation}

The new endcap KLM detector geometry has been coded into the GEANT4
package~\cite{geant}, which then simulates the energy deposition of
signal hits in each strip. Both signal events and neutron background
were simulated in this way. For the neutron background, a $\pm$6~m
accelerator tunnel side was simulated, including a realistic
description of the SuperKEKB optics and also including radiative
Bhabhas, the Touschek effect, and other background sources. The
response of the KLM system to the energy deposition is simulated based
on our test measurements. We assume that multiple hits at the same
strip are correctly resolved by the readout electronics if their
arrival time difference is greater than 10~ns, otherwise they are
merged into one hit. Hit information is stored for further analysis if
the amplitude exceeds the threshold of 7.5~p.e.

The background induced by neutrons turned out to be slightly lower
than those obtained from our conservative extrapolation described
above, corresponding to a maximum 120~kHz rate for the longest strip. For
our background studies, we assume that hits are distributed uniformly
in time. We use similar procedures to digitize both background and
signal hits.

\subsection{Superlayer cluster reconstruction}

The KLM hit reconstruction procedure is similar to that used in the
Belle reconstruction software. We exploit the improved time resolution
that can be achieved with the new electronics and DAQ~\cite{varner}.

Neighboring hit strips in the same plane with hit time difference
smaller than 10~ns are grouped to form a 1D hit with coordinate
defined as the weighted average over the hit-strips' coordinates. Two
1D hits from orthogonal layers in the same superlayer that intersect
spatially and match in time form a `2D hit'. The reconstructed 2D hits
from all superlayers are used for muon and \kl\ reconstruction.

\subsubsection{Muon identification}

The muon identification procedure, for muons exceeding the detection
threshold of 0.6\gevc , is based on the extrapolation of reconstructed
charged tracks outward from the Belle II drift chamber into the KLM
detector. The muon identification procedure then compares the
likelihoods of muon vs. hadron hypotheses, based on the longitudinal
and transverse profiles of the extrapolated tracks to EKLM 2D hits
within some geometric matching requirement.

Misidentification is mainly due to hadronic showers and pion or kaon
decay in flight, which is thus largely independent of the KLM
detection technique. However, the expected efficiency of the
scintillator EKLM detector is slightly higher than for the previous
Belle EKLM; we therefore expect that the muon identification
performance should be slightly improved.
 
The measured muon detection efficiency and pion fake rates are
approximately constant for momenta greater than 1.0 and 1.5\gevc,
respectively. Between 1.0 and 3.0\gevc\ the average muon detection
efficiency is 90\% and the hadron fake rate is less than 1.5\% over
the KLM acceptance, using the standard likelihood selection criterion
$\mathcal{L}>0.9$.

\subsubsection{\kl\ reconstruction}

\begin{figure}[htb]
\begin{center}
\includegraphics[width=0.92\textwidth]{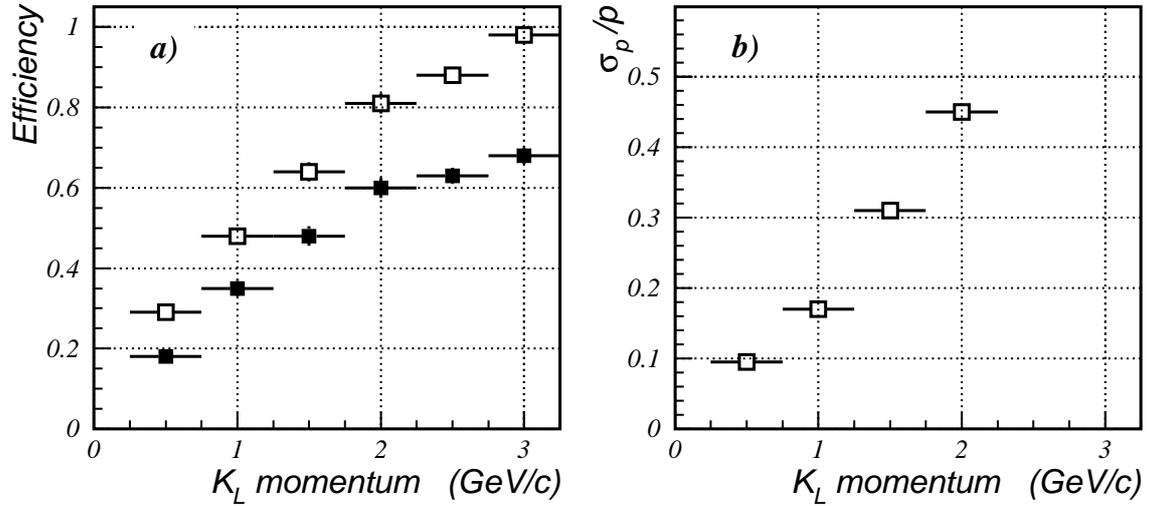} 
\end{center}
\caption{\kl\ reconstruction: a) efficiency of \kl\ cluster-finding
  with a tight requirement of two 2D-hits in different superlayers
  (hatched squares) vs. a loose requirement of $\geq 1$ superlayers
  hit (open squares); b) momentum resolution.}
\label{fig:KL}
\end{figure}

Following the procedure used at the Belle experiment, \kl\ clusters
are reconstructed as a group of 2D hits within a $5^\circ$ cone angle
relative to the interaction point, which are compatible in time with
being produced by a single nuclear shower. As in the algorithm used at
Belle we require 2D hits in at least two superlayers. The efficiency
for \kl\ cluster-finding in the KLM is shown in Fig.~\ref{fig:KL}a as
a function of \kl\ momentum. The angular resolution is $\sim$10~mrad
over a wide momentum range, which is approximately the same as for the
Belle system.

The main source of \kl\ fake candidates is the random coincidence of
1D hits induced by neutrons. After SiPM irradiation, some small
contribution to fake candidates arises from the increased SiPM
noise. We find $\ll 0.01$ fake \kl\ candidates per event at the design
luminosity. Given that the fake rate is negligibly small, we have
studied the possibility of increasing the \kl\ efficiency by loosening
the cluster selection, allowing 2D hits in only one superlayer. This
results in increased \kl\ efficiency by up to $30\%$
(Fig.~\ref{fig:KL}a, open squares), while maintaining the background
rate at a relatively low level ($<0.2$ fake cluster/event). Further
fake cluster rate suppression can exploit both amplitude information,
as well as the one-superlayer hit pattern characteristics, depending
on the signal-to-noise requirements of a particular analysis.

We can also take an advantage of the good scintillator time resolution
to determine the \kl\ momentum using the EKLM system as a
time-of-flight detector, at least for medium-momentum kaons. The
resulting momentum resolution as a function of \kl\ momentum is shown
in Fig.~\ref{fig:KL}b. The timing measurements provide useful
information for kaons with momenta up to 1.5\gevc.

Based on the simulation we conclude that the new EKLM detector
provides even better performance for \kl\ reconstruction than the old
one, and thus also an improved hadronic veto for the study of
important $B$ decays modes with neutrinos in the final state.

\section{Conclusion}

We have studied a system based on scintillator counters with WLS fiber
light collection and SiPM readout for the Belle II experiment. We have
identified a few simple improvements in the strip production
technology which allow significant increases in the light collection
efficiency, thus increasing the efficiency and robustness of the
entire detector. The new system should work efficiently at background
rates and radiation doses $\sim$100 times larger than those observed
for the Belle experiment. As demonstrated by many tests described
herein, the system has sufficient robustness to operate well in a
strong magnetic field and high radiation and interaction environment
with no significant degradation anticipated after many years of
data-taking. While this system was designed for a particular
experiment, namely Belle II, our study can be applied to the
construction of muon systems in many experiments. For example, our
experience is being used in the design of the cosmic ray muon veto
system for the COMET experiment~\cite{Comet}. It was also considered
during the design of the muon system of the ILD experiment at the
International Linear Collider (ILC)~\cite{ILD_TDR}.

\section{Acknowledgments}

It is our pleasure to thank K.\,Abe and K.\,Sumisawa for their
invaluable help with the carrying tests of the module at KEK. The work
is supported by the Russian Ministry of Education and Science
contracts 14.A12.31.0006 and 4465.2014.2 and the Russian Foundation
for Basic Research grant 14-02-01220.


\begin{thebibliography}{9}

\bibitem{LoI} S.\,Hashimoto {\em et al.,} editors, \emph{Letter of
  Intent for KEK Super B Factory (Part-II, Detector)}, KEK Report 04-4
  (2004).

\bibitem{belle2tdr} T.\,Abe {\it et al.}, \emph{The Belle II Technical
  Design Report}, KEK Report 2010-1 (2010), arXiv:1011.0352.

\bibitem{belle} A.\,Abashian {\it et al.} (Belle Collaboration),
  Nucl. Instrum. Methods A {\bf 479}, 117 (2002).

\bibitem{RPC} A.\,Abashian {\it et al.} (Belle Collaboration),
  Nucl. Instrum. Methods A {\bf 449}, 112 (2000).

\bibitem{ITEP_Strip} V.\,Balagura {\em et al.}, Nucl. Instrum. Methods
  A {\bf 564}, 590 (2006).

\bibitem{superb} M.\,Andreotti {\em et al.}, IEEE
  Nucl. Sci. Symp. Conf. Rec. 1718 (2010).

\bibitem{Minos} I.\,Ambats {\em et al.} (Minos Collaboration),
  FERMILAB-DESIGN-1998-02 (1998).

\bibitem{Opera} R.\,Acquafredda {\em et al.} (Opera Collaboration),
  JINST, 4, P04018 (2009) (doi: 10.1088/1748-0221/4/04/P04018).
 	
\bibitem{SiPM} A.\,Akindinov {\em et al.}, Nucl. Instrum. Methods A
  {\bf 387}, 231 (1997); \\ G.\,Bondarenko {\em et al.},
  Nucl. Phys. Proc. Suppl. B {\bf 61}, 347 (1998); \\ V.\,Saveliev and
  V.\,Golovin, Nucl. Instrum. Meth. A {\bf 442}, 223 (2000);
  \\ G.\,Bondarenko {\em et al.}, Nucl. Instrum. Methods A {\bf 442},
  187 (2000); \\ Z.\,Sadygov {\em et al.}, Nucl. Instrum. Methods A
  {\bf 504}, 301 (2003); \\ V.\,Golovin and V.\,Saveliev,
  Nucl. Instrum. Meth. A {\bf 518}, 560 (2004).

\bibitem{Danilov} M.\,Danilov, Nucl. Instrum. Methods A {\bf 581}, 451
  (2007).

\bibitem{Sefkow} F.\,Sefkow, PoS PD07:003 (2006).

\bibitem{T2K} Y.\,Kudenko {\em et al.} (T2K Collaboration),
  Nucl. Instrum. Methods A {\bf 598}, 289 (2009).

\bibitem{hamamatsu} Hamamatsu Photonics K.K. (Japan), {\tt
  http://www.hamamatsu.com/eu/en/index.html}

\bibitem{cpta} CPTA, Ltd. (Russia), {\tt
  http://www.cpta-apd.ru/eng/index.html}

\bibitem{mephi} Research and production enterprise PULSAR, {\tt
  http://www.pulsarnpp.ru/index.php/fotoelektronika} {\em (in
  Russian)}.

\bibitem{Andreev} V.\,Andreev {\em et al.}, Nucl. Instrum. and Methods
  A {\bf 540}, 368 (2005).

\bibitem{bicron} Saint-Gobain, Ltd. (France), {\tt
  http://www.crystals.saint-gobain.com/Plastic\_Scintillation.aspx.}

\bibitem{kharkov} AMCRYS-H Co., (Ukraine), {\tt
  http://www.amcrys-h.com/index.htm.}

\bibitem{fermilab} A.\,Pla-Dalmau {\em et al.} (MINOS Scintillator
  Group Collaboration), Frascati Phys. Ser. {\bf 21},513 (2001).

\bibitem{vladimir} Research and production enterprise UNIPLAST,
  (Russia), {\tt http://uniplast-vladimir.com/index.php?page=scint}
  {\em (in Russian)}.

\bibitem{T2K_tdr} P.-A.\,Amaudruza, {\em et al.} (T2K Collaboration),
  Nucl. Instrum. Methods A {\bf 696}, 1 (2012).

\bibitem{Kuraray} Kuraray, Ltd. (Japan), {\tt
  http://kuraraypsf.jp/psf/ws.html}.

\bibitem{surel} Research and production enterprise SUREL,
  Ltd. (Russia), {\tt http://www.surel.ru/silicone/36-surel-sl-1.php}
  {\em (in Russian)}.

\bibitem{Calice} B.\,Dolgoshein {\em et al.}, Nucl. Instrum. Methods A
  {\bf 563}, 368 (2006).

\bibitem{Hera-b} A.\,Zoccoli {\em et al.} (Hera-B Collaboration),
  Nucl. Instrum. Methods A {\bf 446}, 246 (2000), \\ 
    Jr. A. A. Alves {\em et al.} (LHCb Collaboration), JINST, 3:S08005
    (2008).

\bibitem{hardness} see eg. P.\,Perret, arXiv:1407.4289.

\bibitem{Tarkovsky} E.\,Tarkovsky, PoS PD07, 013 (2006).

\bibitem{luxel} Landauer, Ltd. (Japan), {\tt
  http://www.landauer.com/Industry/Products/Dosimeters/Dosimeters.aspx}.

\bibitem{belleii-physics} T.\,Aushev {\em et al.}, arXiv:1002.5012.

\bibitem{geant} S.\,Agostinelli {\em et al.}, (GEANT4 Collaboration),
  Nucl. Instrum. Meth. A {\bf 506}, 250 (2003); J. Allison {\em et
    al.}, (GEANT4 Collaboration), IEEE Transactions on Nuclear
  Science, {\bf 53}, 270 (2006); {\tt http://www.geant4.org/geant4}.

\bibitem{varner} See talk by G.\,Varner: {\tt
  http://www.phys.hawaii.edu/\~idlab/taskAndSchedule/KLM/KLM\_electronics\_CDR.pdf}

\bibitem{Comet} O.\,Markin, E.\,Tarkovsky, arXiv:1402.5522.

\bibitem{ILD_TDR} T.\,Behnke {\em et al.}, arXiv:1306.6329.

\end{thebibliography}
\end{document}